\documentclass[%
 reprint,
showpacs,preprintnumbers,
 amsmath,amssymb,
pra,
]{revtex4-1}

\usepackage{graphicx}
\usepackage{dcolumn}
\usepackage{bm}

\begin{document}


\title{Blind quantum computation with completely classical client and a trusted center}%

\author{Min Liang}
\email{liangmin07@mails.ucas.ac.cn}
\affiliation{Data Communication Science and Technology Research Institute, Beijing 100191, China}


\begin{abstract}
Blind quantum computation (BQC) enables a client without enough quantum power to delegate his quantum computation to a quantum server, while keeping the input data, the algorithm and the result unknown to the server. In the studies of practical BQC protocol, an important problem is how to reduce the quantum requirement of the client.
Multi-server BQC protocols have been proposed to solve this problem. We review the double-server and triple-server protocols [Li et al., Phys. Rev. A 89, 040302(R) (2014)], and propose a modified double-server BQC protocol with a trusted center. In our protocol, the servers are allowed to communicate mutually, and the client is completely classical.
Furthermore, our double-server protocol can be modified into a single-server protocol by simply combining the two servers.
Compared with the triple-server protocol, our double-server and single-server protocols are more simple and the client is not required to have the ability to access quantum channel. So our protocols are more practical when quantum computer is applied in the `cloud' model.
\end{abstract}

\maketitle


\section{Introduction}
Blind quantum computation (BQC) is a kind of secure computation protocol involving two roles, where the client has the data and the algorithm, and the server has a quantum computer. In BQC protocol, the client instructs the server to perform quantum computation on his data without leaking any information about the data, the algorithm and even the final result. BQC protocol should have the properties of blindness, correctness and verifiability. The blindness means that the data, the algorithm and the result are only known by the client; the correctness means that the client can obtain the correct outcome after the protocol being finished; the verifiability means that the client can check whether the server performs the computation following his instructions.

The research of BQC protocol begins from Ref.\cite{childs2005}, which construct some BQC protocols for each of the basic quantum gates (Hadamard gate, CNOT gate, and $\pi/8$ gate). Because any quantum circuit can be implemented by a combination of these basic gates, the BQC protocol can implement any quantum circuit. However, this protocol does not satisfy the requirement of verifiability. Arrighi and Salvail \cite{arrighi2006} proposed another BQC protocol, which is not universal since it is only used for computing some classical functions. Based on quantum authentication scheme, Aharonov et al. \cite{aharonov2008} proposed an interactive proof system for quantum computation, which can also be used as a BQC protocol. However, the client must perform lots of quantum computation in the quantum authentication encoding.
Based on measurement-based quantum computation model, Broadbent et al. \cite{broadbent2009} proposed the first universal BQC protocol, in which the client only needs to prepare some single-qubit states.
Currently, lots of researches have been proposed to make the BQC protocols more optimal or practical \cite{sueki2013,morimae2012a,morimae2010,morimae2012b,dunjko2014,giovannetti2013}.
For example, Ref.\cite{morimae2012a} improved the fault-tolerant threshold for BQC by topological quantum computation, and Ref.\cite{mantri2013} studied the optimization of the BQC protocol from the aspect of quantum communication.

All the mentioned protocols are single-server BQC protocols, in which the client must have some quantum abilities, such as the ability to prepare or measure single qubit \cite{broadbent2009,morimae2013a}. An important effort in the researches of practical BQC protocol is reducing the quantum requirement of the client.
If the client is limited to only be able to carry out classical computation, we have to increase the number of the servers and devise the multi-server BQC protocols. Actually, Broadbent et al.\cite{broadbent2009} proposed a double-server protocol, where the client is completely classical and the two servers cannot communicate mutually. The double-server protocols in Ref.\cite{morimae2010,morimae2013b} also assume the noncommunicating of the two servers. Later, Li et al.\cite{li2014} proposed a triple-server BQC protocol, in which the client is almost classical (the client can only access the quantum channel) and the servers are allowed to communicate with each other.

We review and analyze the main multi-server BQC protocols, and devise a double-server BQC protocol with completely classical client, where the two servers are allowed to communicate mutually. In our protocol, the trusted center should prepare Bell states and send them to the servers. In addition, it is necessary that there exists a private classical channel from the trusted center to the client.

\section{Reviews and analysis}
Before introducing our double-server BQC protocol, we review
the single-server BQC protocol and double-server BQC protocol in Ref.\cite{broadbent2009}, the modified
double-sever BQC protocol in Ref.\cite{morimae2013b}, and the triple-server BQC protocol in Ref.\cite{li2014}.

Let the set $S=\left\{0,\frac{\pi}{4},\frac{2\pi}{4},\ldots,\frac{7\pi}{4}\right\}$, and $\phi\in_r S$ means $\phi$ is randomly chosen from the set $S$. The notation $|\pm\phi\rangle$ denotes the qubit $|0\rangle\pm e^{i\phi}|1\rangle$.

\subsection{BFK single-server protocol}
Client wants to finish a quantum computation. Assume he has in mind the quantum computation on the $n$-qubit graph state corresponding to the graph $G$.
The quantum operation that Client wants to perform is to measure the $i$th qubit in the basis $\{|\pm\phi_i\rangle\}$.
\begin{description}
  \item[S1] Client prepares $n$ qubits and sends them to Server. The state of each qubit is $|\theta_i\rangle=|0\rangle+e^{i\theta_i}|1\rangle(i=1,2,\ldots,n)$, where $\theta_i$ is uniformly chosen from the set $S$.
  \item[S2] Server produces the brickwork state $|G(\theta)\rangle$ by applying controlled-Z gates on the received qubits based on the graph $G$.
  \item[S3] For $i=1,2,\ldots,n$, Client randomly chooses $r_i\in\{0,1\}$ and computes $\delta_i=(\theta_i+\phi'_i+r_i\pi)\textrm{mod}2\pi$, where $\phi'_i$ is obtained according to the previous measurements and $\phi_i$, and then $\delta_i$ is sent to Server if Client needs Server to measure the $i$th qubit of $|G(\theta)\rangle$.
  \item[S4] Server performs a measurement on the $i$th $(i=1,2,\ldots,n)$ qubit in the basis $\{|\pm\delta_i\rangle\}$ and informs Client about the measurement result.
\end{description}

This single-server BQC protocol can be modified to be a
BQC protocol with authentication where a cheating server can
be found with probability exponentially approaching one \cite{broadbent2009}. All the protocols that are introduced in this article do not have the property of authentication, however, they can also be modified to be a BQC protocol with authentication in the similar way. More details
can be found in Ref.\cite{broadbent2009}. We do not discuss the authentication here.

Notice that there is no explicit input in the protocol.
In fact, the preparation of the initial input state can be included in the algorithm of quantum computation.
For example, in the steps {\bf S3} and {\bf S4}, Client can instruct Server to prepare the input state and perform the quantum computation on the state.
Obviously, the input state is still unknown to Server.

\subsection{Double-server BQC protocols}
Broadbent et al.\cite{broadbent2009} point out an important problem: does there exist a BQC protocol which permits a completely classical client?
They present a solution which is based on two entangled servers; However, the two servers are not allowed to communicate with each other.
In the following, we firstly review this double-server BQC protocol \cite{broadbent2009} (it is named BFK double-server protocol in this article), and then introduce a modified version \cite{morimae2013b}.

Bell states are used in the protocols.
The Bell state of a pair of particles $(a,b)$ can be denoted by $|\varphi_{x,z}(a,b)\rangle=(I\otimes X^xZ^z)\frac{1}{\sqrt{2}}(|00\rangle_{a,b}+|11\rangle_{a,b})$,
where $x,z\in\{0,1\}$.

{\bf Initialization: The trusted center distributes Bell states to Server1 and Server2.}

The trusted center prepares $n$ Bell states $|\varphi_{0,0}(A_i,B_i)\rangle$ ($i=1,\ldots,n$).
He sends the $n$ particles $A_1,\ldots,A_n$ to Server1, and sends the $n$ particles $B_1,\ldots,B_n$ to Server2.

{\bf Stage 1: Client's preparation with Server1}
\begin{description}
  \item[D1-1] Client independently chooses $n$ random values $\tilde{\theta}_i\in_r S$ ($i=1,\ldots,n$), and sends them to Server1;
  \item[D1-2] According to the $n$ bases $\{|\pm\tilde{\theta}_i\rangle\}_{i=1}^n$, Server1 performs quantum measurements on his particles $A_1,\ldots,A_n$ and obtains $n$ bits, which are denoted as $\{m_i\}_{i=1}^n$. Then he sends these bits to Client.
\end{description}

{\bf Stage 2: Client's computation with Server2}
\begin{description}
  \item[D2-1] Similar to {\bf S2}, Server2 produces the brickwork state using all his qubits;
  \item[D2-2] Client starts the BFK single-server BQC protocol with Server2 from the step {\bf S3}, replacing $\theta_i$ with $\tilde{\theta}_i+m_i\pi$ ($i=1,\ldots,n$).
\end{description}

In this double-server BQC protocol, Client is only required to do classical computation, and the two servers are not allowed to communicate with each other.
It is worth to notice that, the information $\{\tilde{\theta}_i\}_{i=1}^n$ and $\{m_i\}_{i=1}^n$ should be transmitted through a private channel (for example, a secure encryption algorithm is used) between Client and Server1; Otherwise, Server2 may obtain the values of $\{\theta_i\}_{i=1}^n$ and break the blindness of the protocol.

A modified version of the above double-server BQC protocol is proposed by Morimae and Fujii in Ref.\cite{morimae2013b}. They present a secure entanglement distillation scheme for the double-server BQC protocol. Ref.\cite{sheng2015} gives out a better entanglement distillation scheme with higher efficiency. We briefly review Morimae and Fujii's protocol without entanglement distillation as follows.

{\bf Initialization: The trusted center distributes Bell states $|\varphi_{x_i,z_i}(A_i,B_i)\rangle$ to Server1 and Server2, where $x_i,z_i$ are known by Client and $i\in\{1,\ldots,n\}$.}

{\bf Stage 1: Client's preparation with Server1}
\begin{description}
  \item[M1-1] Client independently chooses $n$ random values $\{\tilde{\theta}_i\in_r S\}_{i=1}^n$, and sends the messages $\{\tilde{\theta}'_i=(-1)^{x_i}\tilde{\theta}_i+z_i\pi\}_{i=1}^n$ to Server1;
  \item[M1-2] According to the $n$ bases $\{|\pm\tilde{\theta}'_i\rangle\}_{i=1}^n$, Server1 performs quantum measurements on his particles $A_1,\ldots,A_n$, and obtains $n$ bits $\{m_i\}_{i=1}^n$; Then he sends these bits to Client.
\end{description}

{\bf Stage 2: Client's computation with Server2}
\begin{description}
  \item[M2-1] Same as {\bf D2-1}.
  \item[M2-2] Same as {\bf D2-2}.
\end{description}

The above protocol is an improved version of BFK double-server BQC protocol. In the BFK double-server BQC protocol, the shared entanglement between the two servers is $\frac{1}{\sqrt{2}}(|00\rangle+|11\rangle)$, and Server1 performs quantum measurement on the first particle according to the basis $|\pm\tilde{\theta}_i\rangle$; In the modified version, the shared entanglement between the two servers is $\left(I\otimes X^{x_i}Z^{z_i}\right)\frac{1}{\sqrt{2}}(|00\rangle+|11\rangle)$,
and Server1 performs quantum measurement on the first particle according to the basis $|\pm((-1)^{x_i}\tilde{\theta}_i+z_i\pi)\rangle$.
Actually, they are equivalent since the measurement outcome of the particle $a$ in $|\varphi_{x_i,z_i}(a,b)\rangle$ according to the basis $|\pm((-1)^{x_i}\tilde{\theta}_i+z_i\pi)\rangle$
is equal to the measurement outcome of the particle $a$ in $|\varphi_{0,0}(a,b)\rangle$ according to the basis $|\pm\tilde{\theta}_i\rangle$.

\subsection{Triple-server BQC protocol}
In the reviewed double-server protocols, the two servers are not allowed to communicate with each other. It is unrealistic to forbid two quantum servers to communicate mutually. So, the researchers try to solve this problem.
Based on entanglement swapping and the above modified protocol, Li et al.\cite{li2014} devised a triple-server BQC protocol in which the three quantum servers can communicate mutually. However, the client is not completely classical, and must have the ability to access the quantum channel. The protocol has the following four stages.

{\bf Initialization: The trusted center distributes Bell states to Server1, Server2 and Client.}
\begin{description}
  \item[T0-1] The trusted center produces $n=(2+\delta)m$ Bell states $|\varphi_{0,0}(A_i,B1_i)\rangle$ ($i=1,\ldots,n$). He sends the $n$ particles $A_1,\ldots,A_n$ to Client, and sends the $n$ particles $B1_1,\ldots,B1_n$ to Server1.
  \item[T0-2] The trusted center produces $n$ Bell states $|\varphi_{0,0}(A'_i,B2_i)\rangle$ ($i=1,\ldots,n$). He sends the $n$ particles $A'_1,\ldots,A'_n$ to Client, and sends the $n$ particles $B2_1,\ldots,B2_n$ to Server2.
\end{description}

{\bf Stage 1: Client's preparation with Server3}
\begin{description}
  \item[T1-1] For each particle $A_k$ or $A'_l$ arriving, Client randomly chooses one of the two choices: (a) discarding it, or (b) transmitting it to Server3 and recording its position.
  \item[T1-2] Server3 may receive $2m$ particles from Client, where the $m$ particles that are entangled with Server1 are denoted as $A_{s_1},\ldots,A_{s_m}$,
      and $m$ particles that are entangled with Server2 are denoted as $A'_{t_1},\ldots,A'_{t_m}$. Server3 performs Bell measurement on the $m$ pairs of particles $\{(A_{s_i},A'_{t_i})\}_{i=1}^m$, and sends the result $\{(x_i,z_i)\}_{i=1}^m$ to Client.
\end{description}

{\bf Stage 2: Client's preparation with Server1}

\begin{description}
  \item[T2-1] Client chooses $n$ values $\{\tilde{\theta}_i\}_{i=1}^n$ and sends the values of $\{\tilde{\theta}'_i=(-1)^{x_i}\tilde{\theta}_i+z_i\pi\}_{i=1}^n$ to Server1. $\{\tilde{\theta}'_i\}_{i=1}^n$ are distributed as uniformly as possible over all the eight elements of the set $S$.
  \item[T2-2] According to the bases $\{|\pm\tilde{\theta}'_i\rangle\}_{i=1}^n$, Server1 performs quantum measurements on the $n$ particles $B1_1,\ldots,B1_n$, and sends the result $\{m_i\}_{i=1}^n$ to Client. Client only keeps the values of $\{m_{s_i}\}_{i=1}^m$.
\end{description}

{\bf Stage 3: Client's computation with Server2}

\begin{description}
  \item[T3-1] Client asks Server2 to keep the $m$ qubits which are labeled as $\{t_i\}_{i=1}^m$. Notice that the $m$ qubits can be represented as $\{|\tilde{\theta}_{s_i}+m_{s_i}\pi\rangle\}_{i=1}^m$ at this time.
      Similar to {\bf S2}, Server2 produces the brickwork state using the $m$ qubits.
  \item[T3-2] Client starts the reviewed single-server BQC protocol with Server2 from the step {\bf S3}, replacing $\theta_i$ with $\tilde{\theta}_{s_i}+m_{s_i}\pi$ ($i=1,\ldots,m$).
\end{description}

Compared with Morimae and Fujii's double-server BQC protocol \cite{morimae2013b}, the above protocol has added another quantum server Server3. Actually, for the received particles, he just performs some Bell measurements on these particles. If the $i$th measurement result is $(x_i,z_i)$, Client can know the entanglement has been established between Server1's particle $B1_{s_i}$ and Server2's particle $B2_{t_i}$, and the combined state of $B1_{s_i}$ and $B2_{t_i}$ is
$|\varphi_{x_i,z_i}(B1_{s_i},B2_{t_i})\rangle$. Then, Client starts the interactive procedure with Server1 and Server2 that are similar to Morimae and Fujii's double-server protocol.

In the triple-server protocol, the three servers are allowed to communicate mutually, and can be modified to be a single-server BQC protocol, in which the client is also an almost classical user. More details can be found in Ref.\cite{xu2014}. Recently, Ref.\cite{hung2015} points out a security loophole in both this triple-server protocol and the single-server protocol in Ref.\cite{xu2014}

\section{Blind quantum computation with completely classical client}
Generally, single-server BQC protocols require the client to have some quantum ability, such as the ability to produce single-qubit states or make quantum measurements.
If the client can only perform classical computation, the BQC protocols require at least two quantum servers. For example, the double-server BQC protocol in Ref.\cite{broadbent2009} only requires the client to have a classical computer. However, all the double-server protocols \cite{broadbent2009,morimae2010,morimae2013b} assume the servers cannot communicate mutually.
Ref.\cite{li2014} proposes a triple-server BQC protocl, in which the three servers are allowed to communicate mutually. However, the client is almost classical because he needs to access quantum channel.

\begin{figure}[htb!]
\centering
\includegraphics[scale=0.7]{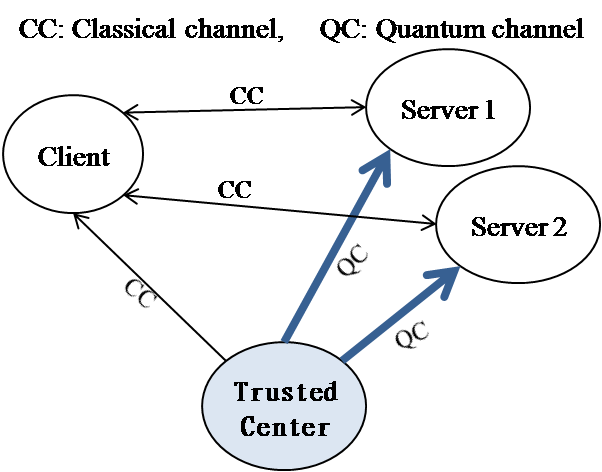}
\caption{The diagram of our modified double-server BQC protocol. In the diagram, `CC' and `QC' represent the classical channel and the quantum channel, respectively. The trusted center and each server are connected by an unidirectional quantum channel. The trusted center and the client are connected by an unidirectional classical channel. The client and each server are connected by the bidirectional classical channel.}
\label{fig1}
\end{figure}

In this section, we will propose a new modified double-server BQC protocol, which allows a completely classical client and the two servers are allowed to communicate mutually. In our protocol, the client and the servers are connected by the bidirectional classical channels, and there exists an unidirectional classical channel from the trusted center to the client. There are an unidirectional quantum channel from the trusted center to each server. See Figure.\ref{fig1}.

The protocol has the following three stages: Initialization, Stage 1 and Stage 2.

{\bf Initialization: The trusted center distributes Bell states to Server1 and Server2, and sends classical secret information to Client.}
\begin{description}
  \item[P0-1] The trusted center randomly chooses $n$ pairs of bits $\{(x_i,z_i)\}_{i=1}^n$ and a $n$-ary permutation $P$, and sends them to Client through the classical channel in a secure way;
  \item[P0-2] The trusted center prepares $n$ Bell states $|\varphi_{x_i,z_i}(A_i,B_i)\rangle$, $i=1,...,n$, and sends the particles $A_1,\ldots,A_n$ to Server1 in sequence. Then he sends the particles $B_{P^{-1}(1)},\ldots,B_{P^{-1}(n)}$ to Server2 in the order which depends on the permutation $P$.
\end{description}

{\bf Stage 1: Client's preparation with Server1}
\begin{description}
  \item[P1-1] Client independently chooses $n$ random values
      $\{\tilde{\theta}_i\}_{i=1}^n$ from the set $S$, and sends the values of  $\{\tilde{\theta}'_i=(-1)^{x_i}\tilde{\theta}_i+z_i\pi~\mathrm{mod}2\pi\}_{i=1}^n$ to Server1;
  \item[P1-2] According to the $n$ bases $\{|\pm\tilde{\theta}'_i\rangle\}_{i=1}^n$, Server1 performs quantum measurements on the $n$ particles $\{A_i\}_{i=1}^n$ and obtains $n$ bits. Denotes the $n$ bits as $\{m_i\}_{i=1}^n$. He sends these bits to Client.
\end{description}

{\bf Stage 2: Client's computation with Server2}
\begin{description}
  \item[P2-1] Similar to {\bf S2}, Server2 produces the brickwork state using the $n$ particles $B_{P^{-1}(1)},\ldots,B_{P^{-1}(n)}$.
  \item[P2-2] Client starts the BFK single-server BQC protocol with Server2 from the step {\bf S3}, replacing $\theta_i$ with $\tilde{\theta}_{P^{-1}(i)}+m_{P^{-1}(i)}\pi$ (or let $\theta_{P(i)}=\tilde{\theta}_i+m_i\pi$), $i=1,\ldots,n$.
\end{description}

According to the step \textbf{P2-2}, it is obvious that Client can run the correct single-server blind quantum computation with Server2.

Next, we show that the above modified double-server BQC protocol is secure even if the two servers communicate mutually.

Suppose Server1 and Server2 can communicate with each other, they may cooperate and attempt to obtain the information related to Client's quantum computation, such as something about $\{\theta_i\}_{i=1}^n$ or $\{\phi_i\}_{i=1}^n$.
Assume that Server2 who knows $\{\delta_i\}_{i=1}^n$ is chosen to do such thing. Server1 tells his information $\{\tilde{\theta}'_i\}_{i=1}^n$ and $\{m_i\}_{i=1}^n$ to Server2. Server2 still cannot learn anything about $\{\theta_i\}_{i=1}^n$ or $\{\phi_i\}_{i=1}^n$. The analysis is as follows.
\begin{enumerate}
  \item Server2 cannot know the values of $\{(x_i,z_i)\}_{i=1}^n$ and the permutation $P$ since they are Client's secret information which are transmitted from the trusted center through a secure channel. Though the Bell measurement on the pair of particles $(A_i,B_i)$ can result the values of $x_i$ and $z_i$, Server2 cannot know which particle is $B_i$ since the order of the particles $\{B_i\}_{i=1}^n$ has been rearranged by the unknown permutation $P$.
  \item Without the knowledge of $\{(x_i,z_i)\}_{i=1}^n$, it is impossible for Server2 to compute the values of $\{\tilde{\theta}_i\}_{i=1}^n$ from $\{\tilde{\theta}'_i\}_{i=1}^n$.
  \item Even if Server2 had gained the information about $\{\tilde{\theta}_i\}_{i=1}^n$, he still cannot know the values of $\{\theta_i\}_{i=1}^n$ since he does not know the random permutation $P$.
  \item From Server2's information $\{\delta_i\}_{i=1}^n$, he cannot obtain any information about $\{\phi_i\}_{i=1}^n$ without the knowledge of $\{\theta_i\}_{i=1}^n$.
\end{enumerate}

Compared with the case that the servers communicate mutually, less information can be known by each server when the servers do not communicate mutually. Thus, our protocol is also secure if the two servers do not communicate with each other.
Notice that the reviewed BFK double-server BQC protocol \cite{broadbent2009} is just a special case of our protocol: let all of $\{(x_i,z_i)\}_{i=1}^n$ be zeroes, and $P$ be the identity permutation.

According to Ref.\cite{hung2015}, the triple-server protocol \cite{li2014} suffers from two kinds of attacks, which can gain the positions $\{s_i, t_i\}_{i=1}^m$. The fundamental cause lies in that the positions are determined during the interactive process. However, in our protocol, the positions are decided by the permutation $P$, which has been chosen randomly before the interactive process. The two attacks are ineffective for our protocol.

In our double-server BQC protocol, the servers are allowed to communicate mutually. If the two servers are joined together (the steps of Server1 and Server2 are finished by one server), we can get a single-server BQC protocol.
It is obvious that the single-server protocol is as secure as our double-server protocol.

Compared with the triple-server protocol \cite{li2014}, the client here is completely classical; However, the client in the triple-server protocol must have the ability to access quantum channel, such as the ability to receive and forward the qubits.

\section{Discussions}
In our protocol, there needs a secure classical channel between the trusted center and the client. This channel can be ensured by certain cryptosystem (block cipher or public-key encryption) in modern cryptography. Moreover, it can also be ensured by `QKD+OTP' (quantum key distribution \cite{bennett1984} plus one-time pad).
If the latter is chosen to protect the classical channel, the client is required to finish the procedure of QKD.
We should stress that, our BQC protocol itself does not require the client's quantum ability.

In our protocol, the client is completely classical, then its output is also completely classical.
It is worth to notice the following two points. Firstly, the input of the computation can be a quantum state. Though the client cannot send quantum input to the servers directly, his quantum input can be generated during the process of quantum computation. Secondly, the client delegates the quantum computation to the quantum servers, so the computation is a quantum algorithm. For example, the client cannot perform Shor's quantum algorithm \cite{shor1994}, but he can finish the algorithm through the interaction with the quantum servers.

Our protocols (double-server protocol and single-server protocol) are very practical while being applied in the cloud quantum computation. In our protocols, the client is completely classical and no quantum channel is needed between the client and the servers. So the widely used classical computers and communication network can satisfy the client's requirement in the implementation of cloud quantum computation. That means, in order to implement cloud quantum computation, the user (who acts as the client) is not required to buy new equipments. We only need to manufacture one (or two) universal quantum computer(s) being used as the server (or Server1, Server2), and set up a trusted center which can prepare Bell states.

The role of the trusted center is very similar to the role of certificate authority (CA) in public-key infrastructure (PKI) \cite{chokhani1994}. In our protocols, the trusted center is needed to prepare Bell states and change the positions of the particles. In addition, a private classical channel is necessary between the client and the trusted center. Compared with the triple-server protocol \cite{li2014}, the trusted center in our protocols is more powerful. However, it is realizable using current quantum technologies.

\section{Conclusions}
This paper reviews the BFK single-server and double-server BQC protocols, the modified double-server protocol in Ref.\cite{morimae2013b} and the triple-server protocol. In these double-server protocols, the client is completely classical, but the servers are not allowed to communicate mutually. In the triple-server protocol, the servers are allowed to communicate mutually, but the cost is higher and the client must have the ability to access quantum channel.
Then we modify the double-server BQC protocols and get a more practical protocol, in which the client is completely classical and the servers are allowed to communicate mutually. We also point out that the double-server protocol can be easily changed to be a single-server protocol without any loss of security.



\begin{thebibliography}{99}
\bibitem{childs2005}
A. M. Childs, Quantum Inf. Comput. 5, 456 (2005).

\bibitem{arrighi2006}
P. Arrighi and L. Salvail, Int. J. Quantum Inform. 4, 883 (2006).

\bibitem{aharonov2008}
D. Aharonov, M. Ben-Or, and E. Eban, {\it Proceedings of the
First Symposium on Innovations in Computer Science} (Tsinghua
University Press, Beijing, 2010), pp. 453 - 469.

\bibitem{broadbent2009}
A. Broadbent, J. Fitzsimons, and E. Kashefi, {\it Proceedings of
the 50th Annual IEEE Symposium on Foundations of Computer
Science} (IEEE, Piscataway, 2009), pp. 517 - 526.

\bibitem{sueki2013}
T. Sueki, T. Koshiba, and T. Morimae, Phys. Rev. A 87, 060301(R) (2013).

\bibitem{morimae2012a}
T. Morimae and K. Fujii, Nat. Commun. 3, 1036 (2012).

\bibitem{morimae2010}
T. Morimae, V. Dunjko, and E. Kashefi, arXiv:1009.3486.

\bibitem{morimae2012b}
T. Morimae, Phys. Rev. Lett. 109, 230502 (2012).

\bibitem{dunjko2014}
V. Dunjko, J. F. Fitzsimons, C. Portmann, and R. Renner, {\it 20th International Conference on the Theory and Application of Cryptology and Information Security} (Springer, Taiwan, 2014), pp. 406 - 425.

\bibitem{giovannetti2013}
V. Giovannetti, L. Maccone, T. Morimae, and T. G. Rudolph, Phys. Rev. Lett. 111, 230501 (2013).

\bibitem{mantri2013}
A. Mantri, C. A. Perez-Delgado, and J. F. Fitzsimons, Phys. Rev. Lett. 111, 230502 (2013).

\bibitem{morimae2013a}
T. Morimae and K. Fujii, Phys. Rev. A 87, 050301(R) (2013).

\bibitem{morimae2013b}
T. Morimae and K. Fujii, Phys. Rev. Lett. 111, 020502 (2013).

\bibitem{li2014}
Q. Li, W. H. Chan, C. H. Wu, and Z. H. Wen, Phys. Rev. A 89, 040302(R) (2014).

\bibitem{sheng2015}
Y. B. Sheng and L. Zhou, Sci. Rep. 5, 7815 (2015).

\bibitem{xu2014}
H. R. Xu and B. H. Wang, arXiv: 1410.7054v2.

\bibitem{hung2015}
S. M. Hung and T. Hwang, arXiv: 1508.07478.

\bibitem{bennett1984}
C. H. Bennett and G. Brassard, {\it Proceedings of IEEE International Conference on Computers, Systems and Signal Processing} (IEEE, Bangalore, 1984), pp. 175 - 179.

\bibitem{shor1994}
P. W. Shor, {\it Proceedings of 35th Annual IEEE Symposium on Foundations of Computer Science} (IEEE, Piscataway, 1994), pp. 124 - 134.

\bibitem{chokhani1994}
S. Chokhani, IEEE Commun. Mag. 32, 70 (1994).

\end{thebibliography}
\end{document}